\documentclass[aps,11pt,pra,showpacs,superscriptaddress,preprint]{revtex4-1}
\usepackage{amsmath}
\usepackage{latexsym}
\usepackage{graphics}
\usepackage{graphicx}
\usepackage{slashed}
\DeclareGraphicsExtensions{.eps,.png}

\usepackage{bm}
\usepackage[bookmarks=true,bookmarksopen=true,bookmarksnumbered=true,bookmarksopenlevel=3]{hyperref}

\begin{document}

\title{Phenomenology of excited doubly charged heavy leptons at LHC}

\author{\textsc{S.~Biondini}} 
\affiliation{Dipartimento di Fisica e INFN, Universit\`{a} degli Studi di Perugia, Via A.~Pascoli, I-06123, Perugia, Italy}
\author{\textsc{O.~Panella}}
\affiliation{Istituto Nazionale di Fisica Nucleare, Sezione di Perugia, Via A.~Pascoli, I-06123 Perugia, Italy}
\author{\textsc{G.~Pancheri}}
\affiliation{Laboratori Nazionali di Frascati, INFN, P.O. Box 13, Frascati, I00044, Italy}
\author{\textsc{Y.~N.~Srivastava}$^1$}
\noaffiliation
\author{\textsc{L.~Fan\`{o}}~$^1$}
\noaffiliation

\date{\today}

\begin{abstract}
We consider the production at the LHC of exotic composite leptons of charge $Q=+2e$. Such states are allowed in composite models which contain extended isospin multiplets ($I_{W}=1$ and $I_{W}=3/2$). These doubly charged leptons couple with Standard Model [SM] fermions via gauge interactions, thereby delineating and restricting their  possible decay channels. We discuss the production cross section at the LHC of $L^{++} \, (p p \rightarrow L^{++}, \, \ell^{-})$ and concentrate on the leptonic signature deriving from the cascade decays $ L^{++} \rightarrow W^{+}\ell^{+} \rightarrow \ell^{+}\, \ell^{+}\, \nu_{\ell}$ i.e. $p\,p \rightarrow \ell^{-} \left( \ell^{+}\, \ell^{+} \right) \, \nu_{\ell}$ showing that the invariant mass distribution of the like-sign dilepton has a sharp end point corresponding to excited lepton mass $m^{*}$. We find that the $\sqrt{s}=7$ TeV run is sensitive at the 3-sigma (5-sigma) level to a mass of the order of 600 GeV if $L=10$ $fb^{-1}$ ($L=20$ $fb^{-1}$). The $\sqrt{s}=14$ TeV run can reach a sensitivity at 3-sigma (5-sigma) level up to $m^{*}=1$ TeV for $L=20$ $fb^{-1}$ ($L$=60 $fb^{-1}$).

\end{abstract}


\maketitle

\section{Introduction}
\label{sec_intro}

The Standard Model [SM] of electroweak interactions explains with great accuracy almost all of the available experimental data. Moreover its aim is to describe all  matter and  interactions through few fundamental constituents. It is the ambitious goal of particle physics to understand how Nature works involving only the smallest number of fundamental elements. However, during the last century, the number of fundamental particles grew, and at present we have a scheme with three generations of quarks and leptons, twelve in all. Furthermore, we have to add fundamental gauge bosons. 
It seems that we have three patterns of quarks and leptons organized in growing masses  but sharing all remaining features [charge, weak isospin, color]. Can we explain this proliferation of fermionic states?  A natural explanation for the replication of fermionic generations could be that they are not truly fundamental particles but instead bound states of some unknown constituents. The idea of further level of compositeness has been investigated phenomenologically for quite some time \cite{ELP}, \cite{CaMaSr}. Such further substructure provides the possibility to have a spectrum of fermions with higher masses than the ones implicit in the SM. For this reason, the observation of any such excited quarks and leptons would be an undeniable signal for compositeness.

There are very different composite models in the literature. They all try to explain the observed quantum numbers of quarks and leptons through more fundamental constituents called \textit{preons}. However, no evidence of form factors \cite{REW1}, \cite{REW2} for SM fermions has been observed even at $\sqrt{s} \simeq$ TeV, and there is not yet any direct evidence of preons. For this reason, it is customary to investigate the consequences of such models phenomenologically~\cite{PDG}, using the effective lagrangian formalism: ignoring the real and more fundamental internal dynamics, we study the effects produced macroscopically, i.e. the transition currents, between the excited fermions and those of the SM. Moreover, there are many different scenarios for compositeness, with energy scales spanning $(1\div 100)$ TeV. Untestable models, from the experimental point of view, also exist which suggest that the energy scale for compositeness may be at a unification scale, i.e. $10^{15}$ GeV or even $10^{19}$ GeV involving the effect of quantum gravity \cite{REW1}.

From the phenomenological point of view, so far the production of excited fermions at colliders has concentrated on multiplets of weak isospin $I_{W}=0$ and $I_{W}=1/2$ and all direct searches within the mass reach of the experiments have failed. At DESY~\cite{DESY} as well as at LEP~\cite{LEP}, no evidence of excited leptons was found, setting  at $95\%$ C.L. a bound $\mathcal{O}(200)$ GeV  on the excited lepton masses. The possibility to search excited quarks and leptons at hadron colliders is discussed in~\cite{Baur:1989kv}, where the authors estimated  the production rates of various signatures at the Fermilab Tevatron and at the Cern LHC. The authors of ref.~\cite{EXCERN}, analyze the potential of the CERN hadron collider to search for excited electrons and neutrinos, showing that LHC will be able to set direct constraints on these possible new states exploring masses up to $1\div 2$ TeV. They employ the theoretical framework of gauge mediated effective lagrangians, concentrating on the excited leptons belonging to $I_{W}=0$ and $I_{W}=1/2$. They reexamine the single production of excited electrons ($e^{*}$) and neutrinos ($\nu^{*}$) via the reactions: $p \, p \rightarrow e^{\pm} \, e^{* \pm} \rightarrow e^{+} \, e^{-} \, V$ and $p \, p \rightarrow \nu^{*} \, e^{\pm} \, e^{* \pm} \, \nu \rightarrow e^{\pm} \, \nu \, V$, where $V$ stands for $\gamma, W^{\pm}, Z$.  

The LHC experiments have already produced new interesting results providing the most stringent bounds on the mass of excited quarks and leptons belonging to $I_{W}=0$ and $I_{W}=1/2$ isospin multiplets. In the quark sector ATLAS \cite{AT1} as well CMS \cite{CMS1} have put upper limits on the search for excited fermions in the di-jet final state, excluding excited quarks up to 2 TeV. New limits for excited lepton masses are provided in \cite{CMS2} where these authors concentrate on the production of exotic leptons via four-fermion interaction and the following electroweak decay $\ell^{*} \rightarrow \ell \, \gamma$. Excited lepton masses are excluded at the $95\%$ C.L. below $1.07$ TeV for electrons and $1.09$ TeV for muons, when compositeness scale is considered $\Lambda=M_{\ell^{*}}\ =\ m^*$.

In this work, we emphasize a particular aspect of compositeness: the weak isospin invariance. In this view, proposed in~\cite{YN}, the usual singlet ($I_W=0$) and doublet ($I_W=1/2$) isospin values are extended to include $I_{W}=1$ and $I_{W}=3/2$. Hence, multiplets (triplets and quartets) appear that contain \emph{exotic} doubly charged leptons of charge $Q=+2 e$ and \emph{exotic} excited quark states of charge $Q=+(5/3) e$. These exotic states are expected to generate interesting signatures to be searched for at the LHC since this accelerator can provide sufficient energy to produce such  new hypothetical massive particles. In this work we will concentrate on the doubly charged excited leptons belonging to the $I_W=1$ and $I_W=3/2$ multiplets. A parallel study concentrating on the signatures of the exotic quarks of charge $Q=(5/3)e$ shall be discussed elsewhere~\cite{qstar}. 

Similar new doubly charged leptonic states have been discussed in the literature. Especially so in connection with mechanisms that generate neutrino masses (type II see-saw) \cite{Chua:2010me,Foot:1988aq,Kumericki:2011hf}, and in models of strong electro-weak symmetry breaking~\cite{delAguila:2010es}.
Doubly charged fermions also appear in the context of extended supersymmetric models as doubly charged higgsinos~\cite{demir,dutta,chacko,frank} and in flavor models in warped extra dimensions and in more general models~\cite{cirelli,delaguila}. 

Very recently, the production at LHC of  doubly charged leptons, belonging to a vector-like triplet with $Y=1$ which mixes with the ordinary leptons of the standard model via a Yukawa coupling to the Higgs has been studied in detail~\cite{PLT}.

In our model~\cite{YN} taking up a composite scenario for quarks and leptons, doubly charged excited leptons, labeled with $L^{--}$, exist  which can couple with the SM lepton only through the $W$ gauge boson. The main consequences are the following: (i) we have only one decay channel for $L^{--} \rightarrow W^{-} \, \ell^-$ with the branching ratio $BR=1$;
(ii) we can produce $L^{--}$ resonantly via $2 \rightarrow 2$ processes such as  $q \bar{q}' \rightarrow L^{--} \ell^+$.
We note that, in principle, the previous limits on excited lepton (quarks) masses derived assuming the usual singlet and/or doublet assignment are not valid for our exotic charged leptons (quarks) which belong to the extended multiplets. Respecting the lepton number conservation, only s-channel [annihilation channel] should be considered. The process of production and decay for $L^{--}$ and its antiparticle $L^{++}$ at the parton level is:
\begin{eqnarray}
u \bar{d} \rightarrow L^{++} \, \ell^- &\rightarrow& W^{+} \, \ell^+ \, \ell^-\nonumber\\
\bar{u} d \rightarrow L^{--} \, \ell^+ &\rightarrow& W^{-} \, \ell^- \, \ell^+
\end{eqnarray} 
As the LHC is a proton proton collider the process describing the production of the $L^{++}$ is expected to have  a larger cross section than the process describing the production of $L^{--}$ because the proton contains two valence $u$ quarks and only one valence $d$ quark, whereas the number of antiquarks is assumed to be the same.  However the cross sections of both processes are expected to be relatively small because  at the parton level they both involve a sea quarks distribution function. 

We will consider  only the leptonic decay channels of the W gauge boson, leading to a final state signature which contains  a tri-lepton and missing energy:
\begin{equation}
p \, p \rightarrow \ell^- \, \ell^+ \, \ell^+ \, \nu_{e}
\end{equation}
In particular, our signature contains a like-sign-dilepton (LSD). The above process stands equivalently for the three generations.  In this work, both doubly charged exotic leptons belonging to 
$I_{W}=1$ and $I_{W}=3/2$ are investigated. 

The rest of the paper is organized as follows. Section \ref{sec_model} describes the extension to higher [greater than $1/2$] isospin values of the effective composite models. In Section \ref{sec_production}, we discuss the production cross section for both $L^{++}$ and $L^{--}$, both at the parton level and at the LHC. Section \ref{sec_signal+background} describes the kinematic features of particles in the final state and we especially focus on the invariant mass distribution of the like-sign-dilepton. Moreover we show the relation between the statistical significance and the integrated luminosity giving a detailed prediction of the requested luminosity in order to observe an excited doubly charged lepton of a given mass $m^{*}$. In section \ref{sec_fast+simulation} we discuss a feasibility study of the experimental search of excited doubly charged lepton performing a simulation of particle reconstruction.    Section \ref{sec_conclusions} contains the final discussion and conclusions.

\section{Extended Isospin Model}\label{sec_model}
In the early days of hadronic physics, much progress was made by using strong isospin to discuss the possible patterns of baryon and meson resonances even when quarks and gluons were still unknown. Flavour $SU(2)$ and later $SU(3)$ were important tools in delineating many properties and subsequent classification of mesonic and baryonic states. In the same spirit, just like a great number of strong resonant low energy states ($\mathcal{O}(1)$ GeV)  were found, we may expect something similar in the electroweak interaction, of course at much higher energies. Here, the Higgs vacuum expectation value parameter $v \simeq 238$ GeV ought to play the role of the energy scale for possible fermionic resonances, thereby an expectation of some new physics at $\mathcal{O}(1)$ TeV scale seems natural. With this point of view, weak isospin spectroscopy could reveal some properties of excited fermions without reference to a direct internal dynamics of the building blocks. Hence, we do not aim to explain the origin of three generations, but assume it, and in addition, that lepton and baryon numbers are separately conserved. It is useful to stress that unlike many other schemes~\cite{Contino:2008hi}, the precise structure of the Higgs channel shall play no role in our analysis.   

We begin then with all SM fermions as belonging to isospin doublets or singlets, as usual, i.e. $I_{W}=0$ and $I_{W}=1/2$, and the electroweak bosons having $I_{W}=0$ and $I_{W}=1$. Thus, only fermionic excited states with $I_{W} \leq 3/2$ can arise provided one only uses the light SM fermions and electroweak gauge bosons. In order to compute the production cross section and decays of these excited fermions, we need to define their couplings to light fermions and gauge bosons. The rules are easily derived referring to weak Isospin and Y (hypercharge). Since, all the gauge fields have $Y=0$, excited fermions can only couple to light fermions with the same Y value. Moreover, to satisfy gauge invariance, we need a transition current containing a $\sigma_{\mu \nu}$ term and not a single $\gamma_{\mu}$, i.e. an anomalous magnetic moment type coupling. This automatically provides current conservation.

The doubly charged leptons under consideration in this work belong to the following isospin multiplets ($I_W=1$ and $I_W=3/2$):
\[ L_1 = \left( \begin{array}{c}
L^{0} \\
L^{-} \\
L^{--} \end{array} \right) ,
\qquad
L_{3/2} = \left( \begin{array}{c}
L^{+} \\
L^{0} \\
L^{-} \\
L^{--} \end{array} \right)\]
With similar multiplets for the antiparticles. While referring to the original work in \cite{YN} for a detailed discussion of all couplings and interactions we discuss  here only the main features of the higher multiplets and write down only   the relevant effective lagrangian density.  We recall \cite{YN} that the higher isospin multiplets ($I_W=1,3/2$) contribute only to the iso-vector current and do not contribute to the hyper-charge current. As a result the particles of these higher multiplets interact with the standard model fermions only through the $W$ gauge field. For the exotic doubly charged lepton of the  $I_{W}=1$ triplet the relevant interaction lagrangian  is:
\begin{equation}
\label{lag1}
\mathcal{L}=\frac{gf_{1}}{m^{*}}\left( \bar{L} \, \sigma_{\mu \nu} \, \partial^{\nu} \, W^{\mu} \, \frac{1+\gamma^{5}}{2} \, \ell \right)  + h. c.
\end{equation}
while that of   the doubly charged component of the $I_{W}=3/2$ multiplets reads:
\begin{equation}
\label{lag32}
\mathcal{L}=
 \frac{gf_{3}}{m^{*}}\left( \bar{L}\sigma_{\mu \nu} \, \partial^{\nu} \, W^{\mu} \, \frac{1-\gamma^{5}}{2} \, \ell  \right)  + h.c.
\end{equation}
In the above equations, $m^{*}$ is the excited fermion mass, and $f_{1}, f_{3}$ are dimension-less coupling constants, 
expected to be of order one, whose precise value could  only be fixed within a specific compositeness model.  At this point, we can obtain the vertices needed to calculate the \textit{parton cross sections} and the excited particles \textit{decays}.

In order to perform the needed numerical calculations of the production cross sections and kinematic distributions, we need to implement our model in a parton level generator. We implemented our model through FeynRules~\cite{FeynRules}, a Mathematica~\cite{mathematica} package that generates the Feynman rules of any given quantum field theory model as specified by a particular lagrangian.  FeynRules allows to specify the output in several formats suitable for specific Feynman diagram calculators. Such interfaces are available for CalcHep/CompHEP~\cite{calc1,calc2,calc3}, FeynArts/FormCalc~\cite{feynarts}, MadGraph/MadEvent~\cite{madgraph1,madgraph2} and Sherpa~\cite{sherpa}. This makes it possible to use it for a new model once and then have it available in any of the above programs.  At the time this work was started the available version of MadGraph (version 4) was not compatible with  our model, see Eqs.~(\ref{lag1},\ref{lag32}), whose effective lagrangians contain non-renormalizable operators. It was possible instead to write the FeynRules output in CalcHep format and therefore the higher isospin model of the excited states has been implemented in CalcHEP. 
 We were then able to compare all the analytical results with the output of numerical CalcHEP sessions dealing with  parton cross sections and decay widths, in order to cross-check the new model.

\section{Production and decay of the doubly charged leptons}\label{sec_production}
\begin{figure}[t!]
\includegraphics[scale=0.55]{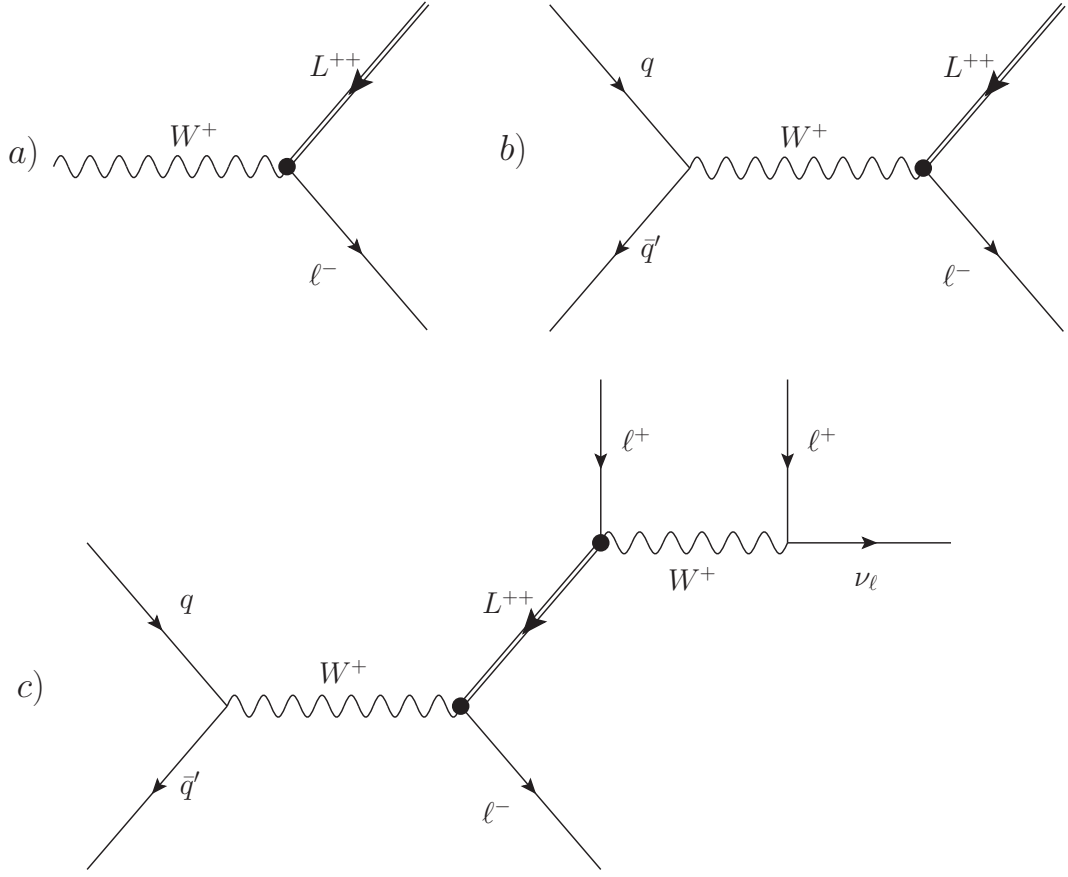}
\caption{\label{fig:fig1} $a)$ The interaction vertex between the exotic doubly charged excited lepton (double line) and the light Standard Model lepton. $L^{++}$ couples only through $W ^{\pm}$ to the ordinary lepton. The black dot denotes the magnetic type coupling. $b)$ The Feynman diagram of the production process for $L^{++}$ at the parton level. $c)$ The only Feynman diagram contributing to the hadron collider process $pp\to \ell^-\ell^+\ell^+\nu_{\ell}$: $L^{++}$ production and  decay with subsequent $W$ gauge boson leptonic decay. }
\end{figure}

\begin{figure}[t!]
\includegraphics[scale=1]{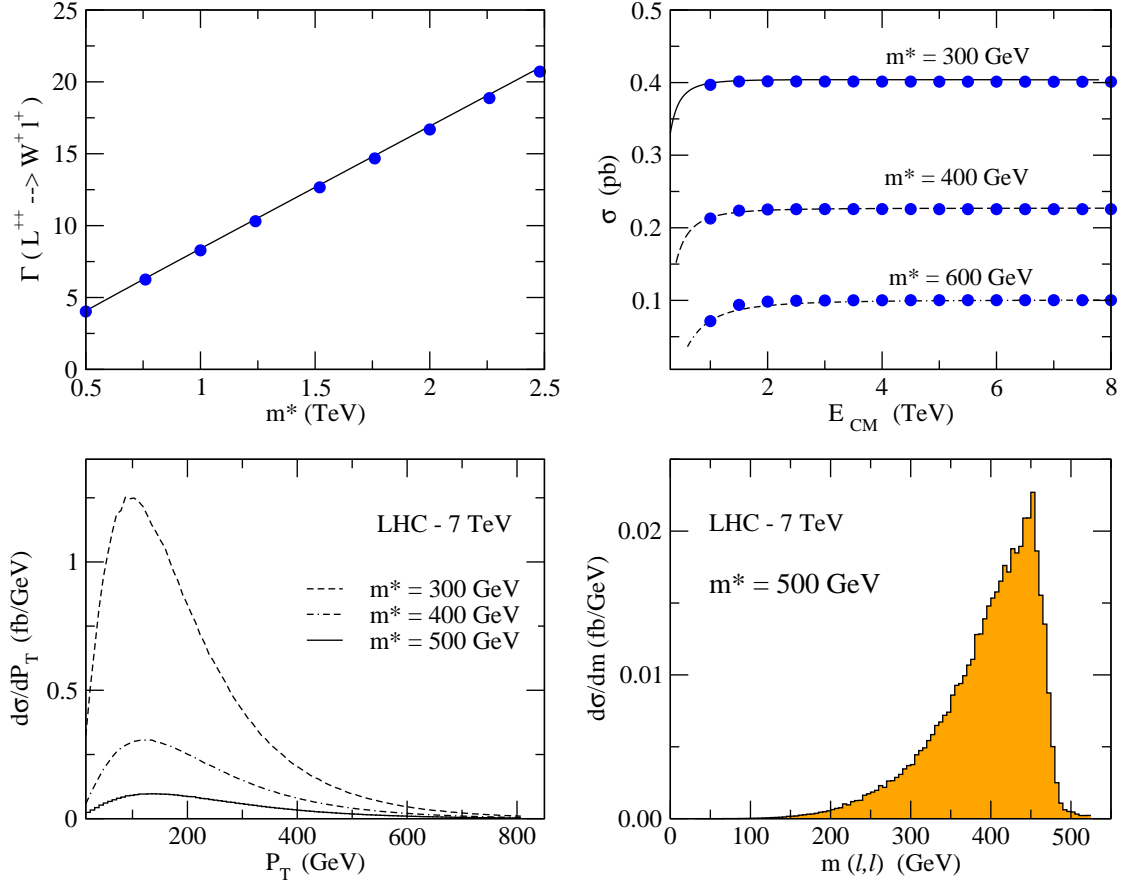}
\caption{\label{parcross} (Color online) Top left: the decay width of the exotic lepton $L^{++}$, as a  function of its mass ($m^*$). The analytical result is compared with the CalcHEP output (red dots); Top right: the total parton cross section for $L^{++}$ or $L^{--}$ as from Eq.~\eqref{SP} against energy in center of mass frame of the partons, for three different values of the excited lepton mass. The analytical results for $m^{*}=300,400,600$ GeV are compared with CalcHEP output (dots);   Bottom left: the transverse momentum distribution of the SM lepton produced in association with $L^{--}; $ Bottom right: invariant mass distribution of the dilepton system, with a sharp end-point at $m_{\ell \ell}\approx m^{*}$. All results have been derived  for a choice of the dimension-less couplings  $f=f_1 = 1$ ($I_W=1$) and/or $f=f_3 = 1$ ($I_W=3/2$) .}
\end{figure}

Considering the interaction lagrangians discussed above  and the fact that  the doubly charged $L^{--}$ and $L^{++}$  interacts with the light fermion only via the $W^{\pm}$ gauge boson, see Fig.~(\ref{fig:fig1}a), one can easily compute the partial and  total decay width of these exotic states.
Indeed the only available decay channel of the doubly charged lepton is $L^{--} \to W^- \ell^-$ with  $\mathcal{B} (L^{--} \to W^- \ell^-)=1$.  The analytic expression  of the total decay width is easily derived:
\begin{equation}
\label{decaywidth}
\Gamma_{L^{++}}= \Gamma (L^{++} \rightarrow W^{+} \, \ell^{+}) = \left( \frac{f}{\sin \theta_{W}} \right)^{2} \alpha_{QED} \frac{m^{*}}{8} \left( 2+ \frac{M^{2}_{W}}{m^{*2}} \right) \left( 1- \frac{M^{2}_{W}}{m^{*2}} \right)^{2} \, ,  
\end{equation}
where $f$ is the dimension-less coupling which depends on the choice of the multiplet: $f=f_1$ for $I_W=1$ and $f=f_3$ when $I_W=3/2$, see Eqs~\eqref{lag1} and Eq.~\eqref{lag32}.
According to the expected large mass for the excited states we can use the approximation  $M_{W} \ll m^{*}$ and Eq.~\ref{decaywidth} suggests that the decay width increases linearly with the mass i.e. $\Gamma= \kappa m^{*} $ as shown in Fig.~\ref{parcross} (top left panel). 


According to lepton number conservation we consider only s-channel for production of doubly charged leptons. The sub-processes are:
\begin{eqnarray}
u \bar{d}&\rightarrow& L^{++} \, \ell^- \; \;  (W^{+} \hbox{-exchange})\\
\bar{u} d &\rightarrow&  L^{--} \, \ell^+ \; \;  (W^{-} \hbox{-exchange})
\end{eqnarray} 
   
Following the convention and notation of \cite{YN} we write down the parton cross section for both $I_{W}=1$ and $I_{W}=3/2$. We give the expressions for $L^{++}$ and its antiparticle using the Mandelstam variables:
\begin{eqnarray}
\left( \frac{d\hat{\sigma}}{d\hat{t}}\right)_{\bar{u}d \rightarrow L^{--} \ell^+}&=&\frac{1}{4m^{*2}s^{2}} \frac{f^2}{12 \pi} \frac{s}{(s-M^{2}_{W})^{2}+(M_{W} \Gamma_{W})^{2}} \nonumber \\ &&\phantom{xxxxxxx}\left\lbrace  (\frac{g^{2}}{4})\left[ m^{*2}(s-m^{*2}) + 2ut \right] \pm 2(-\frac{g^{2}}{8}) m^{*2} (t-u) \right\rbrace  
\label{p1}
\end{eqnarray}
   
\begin{eqnarray}
\left( \frac{d\hat{\sigma}}{d\hat{t}}\right)_{u\bar{d} \rightarrow L^{++} \ell^-}&=&
\frac{1}{4m^{*2}s^{2}} \frac{f^2}{12 \pi} \frac{s}{(s-M^{2}_{W})^{2}+(M_{W} \Gamma_{W})^{2}}\nonumber\\&&\phantom{xxxxxxx} \left\lbrace  (\frac{g^{2}}{4})\left[ m^{*2}(s-m^{*2}) + 2ut \right] \pm 2(-\frac{g^{2}}{8}) m^{*2} (t-u) \right\rbrace  
\label{p2}
\end{eqnarray}
the  $\pm $ refers to $I_W=1,3/2$.
There is a difference which should be stressed. The  $+$ sign in Eq.~\ref{p1} must be used for $I_{W}={3}/{2}$ while the same $+$ sign must be used in Eq.~\ref{p2} for $I_{W}=1$. \textit{We note that if $f_1=f_3$ then at parton level charge conjugation implies the exchange of isospin multiplets}. The expression for the differential parton cross-section reads as following:
\begin{equation}
\label{angulardist}
\frac{d \hat{\sigma}}{ d \Omega}=\frac{g^{4}f^2}{768 \pi m^{*2} \, s} \frac{(s-m^{*2})^{2}}{(s-M^{2}_{W})^{2}+(M_{W} \Gamma_{W})^{2}} \left\lbrace \frac{s}{2} (1-\cos^{2} \theta) +\frac{m^{*2}}{2} (1+\cos^{2} \theta) \pm m^{*2} \cos \theta)\right\rbrace \, ,
\end{equation}
where again $f=f_1$ or $f=f_3$ according to the choice of multiplets, respectively  $I_W=1$ or $I_W=3/2$.
We underline that Eq.~\eqref{angulardist} is valid also for $L^{++}$ production, but as before one should pay attention on the use of the $\pm$ signs as specified above.
Thus we can see a slight difference in the angular distributions between the production of exotic doubly charged leptons belonging to $I_{W}=1$ or $I_{W}=3/2$ isospin multiplets. However such differences in angular distributions disappear when calculating total cross sections at parton level. We obtain for both weak isospin multiplets the following value:
\begin{equation}
\sigma(q \bar{q}' \rightarrow L^{++} \ell^-)=\frac{\alpha^{2}_{QED} f^2}{\sin^{4}\theta_{W}} \frac{V^{qq'}_{CKM}}{36 \pi \, s \, m^{*2}} \frac{(s-m^{*2})^{2}(s+2m^{*2})}{(s-M^{2}_{W})^{2}+(M_{W} \Gamma_{W})^{2}} 
\label{SP}
\end{equation}
Eq.\eqref{SP} is valid  for both $L^{--}$ and its antiparticle (at parton level we have the same production rates).

In Fig.~\ref{parcross} (top right panel) we show sample values of the integrated parton cross section  as function of the partonic center of mass energy for different values ($m^*=300, 400, 600$ GeV) of the excited doubly charged lepton.  We also compare our analytical result with the CalcHEP output as generated with the newly implemented model. The agreement is very good. Fig.~\ref{parcross} (bottom left panel) shows the $p_T$ distribution of the associated standard model lepton in the process $p p \to L^{++} e^-$ which turns out to be peaked for rather hard values of $p_T $.  Finally in the bottom right panel we anticipate the invariant mass distribution of the same sign dilepton in the process $pp\to \ell^+\ell^+ \ell^-  \nu_\ell$ (see Fig.~\ref{fig:fig1}c) to be discussed throughly in section~\ref{sec_signal+background}.

\subsection{Production cross sections}
So far  we have considered the partonic sub-process $u \bar{d} \rightarrow L^{++} \, \ell  $ assuming the quarks to be free particles.  Folding the parton process with the parton distribution functions  (PDFs) the  integrated hadronic cross-section is evaluated as:
\begin{eqnarray}
\sigma=\sum_{a,b} \,\int_{{{m^*}^2}/{s}}^1 \,d\tau \, \frac{d{\cal L}_{a,b}}{d\tau}\, \hat{\sigma}(\tau s, m^*)
\end{eqnarray}
where $\hat{\sigma}(\tau s, m^*)$ is the parton cross section of the sub-process evaluated at the scaled energy $\sqrt{\hat{s}}=\sqrt{\tau s }$ (in the center of mass  frame of the colliding partons) and the differential parton luminosities ${d{\cal L}_{a,b}}/{d\tau}$ are defined as:
\begin{equation}
\frac{d{\cal L}_{a,b}}{d\tau}= \frac{1}{1+\delta_{a,b}}\int_\tau^1 \, \frac{dx}{x}\, \left[f_a(x,\hat{Q}) f_b(\frac{\tau}{x},\hat{Q})+f_b(x,\hat{Q}) f_a(\frac{\tau}{x},\hat{Q})\right]\nonumber
\end{equation}
where in the above equation $\hat{Q}$ is the renormalization and factorization scale at which the parton distribution functions are evaluated.
\begin{figure}[t!]
\includegraphics[scale=1]{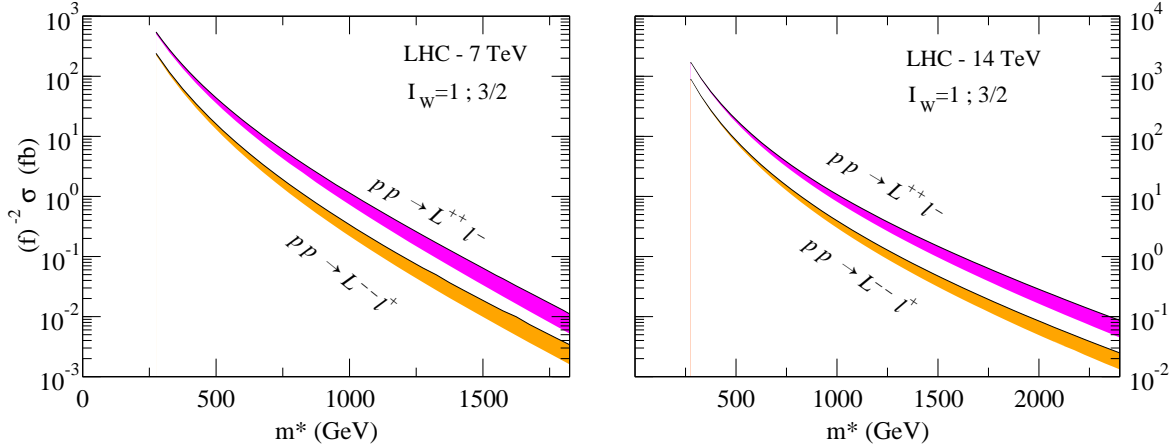}
\caption{\label{prodrates} (Color online) The total integrated cross-sections  ($\sigma$) multiplied by the dimensionless quantity  $f^{-2}$ (where $f$ stands for $f_1$ when $I_W=1$ and $f_3$ when $I_W=3/2$) at LHC energies of $\sqrt{s}=7$ TeV (left panel) and $\sqrt{s}=14$ TeV (right panel) for the production of the exotic leptons $L^{++}$, of charge $Q=+2 e$ and $L^{--}$ of charge $Q=- 2e$.
 We have used  CTEQ6m parton distribution functions~\cite{Pumplin:2002vw}. The uncertainty bands correspond to running the  factorization and renormalization scale from $\hat{Q}=M_W$ (solid line) up to $\hat{Q}=m^*$.
All contributing sub-processes within the first two generations (18) have been summed up. The results are the same for both isospin values $I_{W}=1$ and $I_{W}=3/2$, due to the structure of partonic cross section. See Eqs.~\eqref{p1},\eqref{p2}. }
\end{figure}

Fig.~\ref{prodrates} shows the integrated hadronic cross-sections  $\sigma$ multiplied by the dimensionless quantity  $f^{-2}$ (where $f$ stands for $f_1$ when $I_W=1$ and $f_3$ when $I_W=3/2$)  where we present the results for two different values of the LHC energy, namely $\sqrt{s}=7,14$ TeV. 

We emphasize that the numerical values given in Fig.~\ref{prodrates} refer to the production rates corresponding to $f_1=1$, see Eq.~\eqref{lag1}, and $f_3=1$, see Eq.~\eqref{lag32}.  If other values of such dimensionless couplings should be used the cross sections of Fig.~\ref{prodrates} have to be scaled accordingly, i.e. by multiplying them by $(f_{1})^2$ or $(f_{3})^2$ depending on the iso-spin value .

In the left panel of Fig.~\ref{prodrates}
we show  the total integrated cross section for the production of $L^{++}$ of charge $Q=+2e$ and ${L^{--}}$ of charge $Q=-2e$ for $\sqrt{s}=7$ TeV.   As expected one finds that the production of $L^{++}$ is  larger. This is almost entirely due to the fact that producing $L^{++}$ involves the (dominant) subprocess $u \bar{d} \to L^{++} \ell^-$ i.e. with the valence $u$-quarks in the initial state. On the other end production of the $L^{--}$ state proceeds via the (dominant) subprocess $\bar{u} d \to L^{--} \ell^+$, i.e. with the valence $d$-quark in the initial state. Indeed we see from Fig.~\ref{prodrates} that the $L^{++}$ cross section  is about four times larger than that for $L^{--}$ production. Similar considerations apply to the results at higher energies (right panel). 
 We have used the CTEQ6m parton distribution functions~\cite{Pumplin:2002vw}. The uncertainty bands in 
 Fig.~\ref{prodrates} correspond to running the  factorization and renormalization scale from $\hat{Q}=M_W$ (solid line) up to $\hat{Q}=m^*$.
All contributing sub-processes within the first two generations (18) have been summed up. The results are the same for both isospin values $I_{W}=1$ and $I_{W}=3/2$, due to the structure of partonic cross section. 
See Eqs.~\eqref{p1},\eqref{p2} where the cross section is expressed in terms of the Mandelstam variables.

\section{Other Production Mechanisms}

Before to study in detail our specific like sign dilepton signal and its standard model background we feel that we should briefly account for other mechanisms of production of the exotic heavy fermions object of this work.
In the previous section we have discussed and computed the production of the heavy doubly charged lepton via the magnetic type  transition gauge coupling to the standard model leptons given by Eqs.~(\ref{lag1},\ref{lag32}). 

It is however well known \cite{Baur:1989kv,willenbrock,barger} that at hadron colliders heavy leptons can be pair produced via the standard Drell-Yan mechanism, and also via contact interactions.

The heavy 	quarks and leptons of our model of course will interact at full gauge strength with the Standard Model gauge bosons. The coupling are fixed since we have fixed the quantum numbers (weak isospin and weak hypercharge). We can thus compute for example the couplings to the $Z$ boson. We need only make some hypothesis on the $SU(2)_L\otimes U(1)_Y$ structure of the multiplets. We will make the assumption of a \emph{sequential type} structure with respect to that of the ordinary quarks and leptons:  while the left-handed components are grouped in the $I_W=1, 3/2$  multiplets, the right handed components are singlet with respect to $SU(2)_L$. With this choice we have computed the couplings to the $Z$ boson and in Table~\ref{tabella_COUPLINGS} we report the axial and vector gauge couplings of the $I_W=1$ triplet.
Heavy lepton Drell-Yan production rates  can be computed using the gauge couplings of the new exotic particles given in Table~\ref{tabella_COUPLINGS} from the following formula~\cite{frampton,boyce}:
\begin{eqnarray}
\sigma(q\bar{q} \to L^{++} L^{--}) &=&\frac{2\pi \alpha_{QED}^2}{9\hat{s}} \beta_L \, \left\{(3-\beta_L^2) \, \left[e_q^2 - \frac{2\,e_q \,\, V_q\, V_L}{\sin^2\theta_W\cos^2\theta_W}\frac{\hat{s}(\hat{s}-M_Z^2)}{(\hat{s}-M_Z^2)^2+(\Gamma_ZM_Z)^2}\right]\right.\nonumber\\
&&\left.\phantom{xxxx}+ \frac{(V_L^2+A_L^2)}{\sin^4\theta_W\cos^4\theta_W} \frac{\hat{s}^2 [(3-\beta_L^2)\,(V_q^2)+2\beta_L^2\,A_q^2]}{(\hat{s}-M_Z^2)^2+(\Gamma_ZM_Z)^2}\right\}
\end{eqnarray}
where $\beta_L=\sqrt{1-4m_*^2/\hat{s}}$.
\begin{table}[t!]
\caption{\label{tabella_COUPLINGS}The vector and axial couplings to the $Z$ gauge boson of the components of the $I_W=1$ triplet.}
\begin{ruledtabular}
\begin{tabular}{lccccc}
\phantom{X}& $I_3$&$ C_L$ & $C_R$& $V=(C_L+C_R)/2$ & $A=(C_R-C_L)/2$\cr
\hline
$E^0$& $+1$& $ 1$& $0 $& $ 1/2$ & $-1/2$\cr
\hline
$E^-$& $\phantom{+}0$& $ \sin^2\theta_W$& $ \sin^2\theta_W $& $ \,\sin^2\theta_W$ & $0$\cr
\hline
$E^{--}$& $-1$& $ -1+2\, \sin^2\theta_W$& $ 2\, \sin^2\theta_W $& $ (-1 +4 \,\sin^2\theta_W)/2$ & $+1/2$
\end{tabular}
\end{ruledtabular}
\end{table}  

Another mechanism that has been discussed in the literature with respect to  heavy lepton pair production is  
gluon gluon fusion. This mechanism involves the effective  gluon gluon coupling to the $Z$ gauge boson and Higgs boson arising from exchange at one loop of (heavy) quarks. The computation of the effective $ggZ $  and $ggH$ couplings  is related to the well known triangle anomaly~\cite{bell}.  
In our model  gluon fusion receives no  contribution from the Higgs since our heavy states have no direct coupling to the Higgs as they are expected to acquire their masses through some unknown  mechanism governed by the fundamental preon dynamics.
Following  the notation of \cite{boyce,liu} the gluon fusion cross section is given by:  
\begin{equation}
\label{ggf}
\sigma_{gg}= \frac{(\alpha_{QED}\alpha_s)^2}{128 \pi \sin^2\theta_W} \, \frac{m_*^2}{M_W^4}\, \beta_L \, 
\left|  A_L \frac{\hat{s}-M_Z^2}{\hat{s}-M_Z^2+iM_Z\Gamma_Z}\, \sum_Q A_Q  \left[1+2\lambda_Q I_Q (\lambda_Q)\right]\right|^2
\end{equation}
where $A_L, A_Q$ are the axial couplings to the $Z$ gauge bosons of the heavy leptons and of the quarks in the loop  and $\lambda_Q=m_Q^2/\hat{s}$.

It is well known \cite{willenbrock,liu,boyce,barger} that those quark generations with a high mass splitting between the  highest and lowest isospin components enhance the gluon fusion mechanism. Thus in our model we need only consider the contribution of the $t,b$ quarks since  for the exotic excited quarks  we expect a near degeneracy $m_Q \approx m_L \sim m^*$,  at least to a first approximation. 

It has also been~\cite{Baur:1989kv} noted that assuming electrons and quarks to share common constituents at a hadron collider heavy excited leptons can also be copiously (singly or in pairs) produced via contact interactions.  The effective four-fermion lagrangian  describing, at energies below the compositeness scale, the coupling of excited fermions  to ordinary quarks and leptons  resulting from the strong preon dynamics is:
\begin{equation}
{\cal{L}}_{contact}= \frac{g_*^2}{2\Lambda_C^2} J^\mu J_\mu
\end{equation}
with:
\begin{equation}
J_\mu= \eta_L\, \bar{f}_L\gamma_\mu f_L +{\eta'_L}\, \bar{f}^*_L\gamma_\mu f_L +{\eta''_L}\, \bar{f}^*_L\gamma_\mu f^*_L + h.c. + (L \leftrightarrow R )
\end{equation}
where $g_*^2$ is conventionally chosen equal to $4 \pi$ and left the handed factors $\eta_L$'s are chosen equal to one while the corresponding $\eta_R$'s are chosen to be zero.
The pair production parton cross section via contact interactions is~\cite{Baur:1989kv}: 
\begin{equation}
\hat{\sigma} (q\bar{q}\to L^*\bar{L}^*) = \frac{\pi \beta_L}{12 \hat{s}} \left[ \frac{\hat{s}}{\Lambda_C^2}\right]^2 \left(1+\frac{\beta_L^2}{3}\right)
\end{equation}
and applies equally well to the production of the exotic doubly charged leptons as it involves flavour diagonal contact interactions.
Ref.~\cite{Baur:1989kv} provides a detailed analysis of single and  pair production of excited (but not exotic) leptons and the main conclusion there is that contact interactions provide an important mechanism for producing excited leptons at a high energy hadron collider. Also, contact interactions are responsible for modifying substantially the width of the excited states~\cite{Baur:1989kv}.
We observe that there are now available new strong lower bounds on the value of the contact interaction compositeness scale $\Lambda_{C}$. The ATLAS collaboration has recently reported~\cite{Aad:2011tq} a new analysis of dilepton events from pp collisions at $7$ TeV and a quite stringent  lower bound on the compositeness scale is  derived:  $\Lambda_{C} > 10$ TeV.    This has been obtained by looking for deviations in the production of Drell-Yan pairs ($\mu^+\mu^-, e^+e^-$ dileptons) in $q\bar{q}$ interactions. 

\begin{figure}[t]
\includegraphics[scale=1.0]{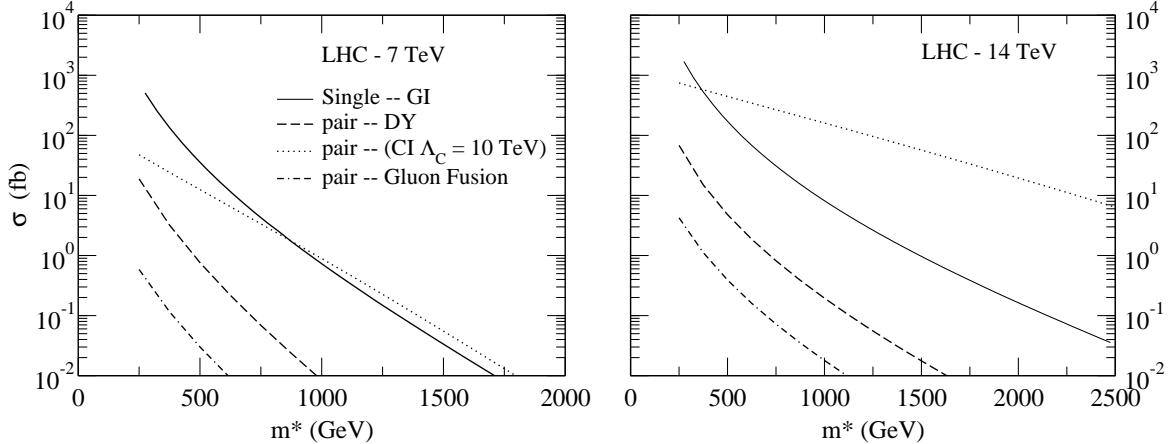}
\caption{\label{mechanisms} Comparison of the various production mechanisms discussed in  section~\ref{mechanisms}. Single production of the exotic doubly charged lepton via gauge interactions  (solid line); Drell-yan pair production (dashed line); pair production via gluon fusion (double dashed-dotted line); pair production via contact interactions with $\Lambda_C = 10$ TeV (dotted line); On the left panel we have the production rates at $\sqrt{s}=7 $ TeV. On the right panel we have the production rates at $\sqrt{s}=10 $ TeV.   We have used  CTEQ6m parton distribution functions~\cite{Pumplin:2002vw} with  the  factorization and renormalization scale fixed at  $\hat{Q}=m^*$. }
\end{figure}

In Fig.~\ref{mechanisms} we compare the various production mechanisms. We clearly see that pair production via the Drell-Yan and gluon fusion is far below the single production via gauge interactions. The gluon fusion mechanism could be somewhat modified (increased) by requiring  that the degeneracy of the exotic quark components is broken. This would require the introduction of at least another parameter (a mass splitting) in the model and we decide not to consider this possibility any further in this work.

On the other hand production of excited  leptons via contact interactions is seen to be rather important. In Fig.~\ref{mechanisms} the dotted curve is the pair production of the  exotic lepton $L^{++}$ via flavor conserving and diagonal~\cite{Baur:1989kv,Eichten:1984eu} contact interaction terms  parametrized with $\Lambda_C=10$ TeV ($\approx$ the current ATLAS bound). We have checked that our numerical results coincide with those of \cite{Baur:1989kv} if one were to rescale appropriately $\Lambda_C$.  

We see that at $\sqrt{s}=7 $ TeV pair production of the exotic doubly charged leptons via contact interactions is already at the same level of the single production via gauge interactions. At $\sqrt{s}=14 $ TeV it will be even dominant (unless the lower bound on $\Lambda_C$ is further increased).

Let us conclude this section with a final remark. One could also consider the contribution of possible flavor conserving~\cite{Baur:1989kv} but \emph{non-diagonal} contact interactions~\cite{Eichten:1984eu,Abolins:1982vy} which  could trigger  the single production of the exotic doubly charged leptons  ($u\bar{d} \to L^{++}\ell^-$) object of this study. However, these type of interactions rely on the further model dependent assumption that  also quarks of different flavor would have to  share common constituents, (in addition to quarks and leptons). 
While this would be without any doubt an interesting possibility we decided to concentrate, in this work, on the pure gauge model of magnetic type transition couplings demanding a complete analysis of the interplay between flavor non-diagonal contact interactions and the pure gauge model to a future work.

\section{Signal and background}
\label{sec_signal+background}
Let us now go back to a detailed discussion  of the kinematic features of the signature $p \, p \rightarrow \ell^- \, \ell^+ \, \ell^+ \, \nu_{e}$ arising from the decay chain of the $L^{++}$ produced via the magnetic type gauge interactions.   
We finally want to show the same sign dilepton (SSDL) invariant mass distribution $m_{(\ell^+,\ell^+)}$   defined by $m^2_{(\ell^+,\ell^+)} = ( p_{\ell^+_{1}} +  p_{\ell^+_{2}} )^2$.   


We analyze the particle set of the final state, which is characterized by the experimental signature $\ell^- \, (\ell^+ \, \ell^+)_{SSDL} \, + \slashed{E}_{T}$. Now we are considering particles coming from two decays, with the topology shown in Fig.~\ref{fig:fig1}, which is often called dilepton topology. In this case a good method for the reconstruction of the mass of the resonant  particle is represented from the following technique. As suggested in~\cite{MR} one should observe a like sign dilepton invariant mass distribution with a sharp end point (jacobian peak) at $m^{*}$, which is also close to the maximum of the distribution. From kinematics one has:
\begin{equation}
\left[m^{2}_{(\ell^+,\ell^+)}\right]_{\text{max}}=\frac{\left( m^{*2}-m^{2}_{W}\right)\left( m_{W}^{2}-m^{2}_{\nu} \right)  }{m^{2}_{W}} \, \rightarrow \, \left[m^{2}_{(\ell^+,\ell^+)}\right]_{\text{max}} = m^{*2}-m^{2}_{W}
\end{equation}
According to the values of excited lepton mass considered in our work the condition $m^{*} \gg m_{W}$ is generally satisfied, so $m_{(\ell^+,\ell^+)} \simeq m^{*}$.
We have obtained an invariant  mass distribution for the SSDL system by CALCHEP numerical sessions. 

\begin{figure}[t]
\includegraphics[scale=1.1]{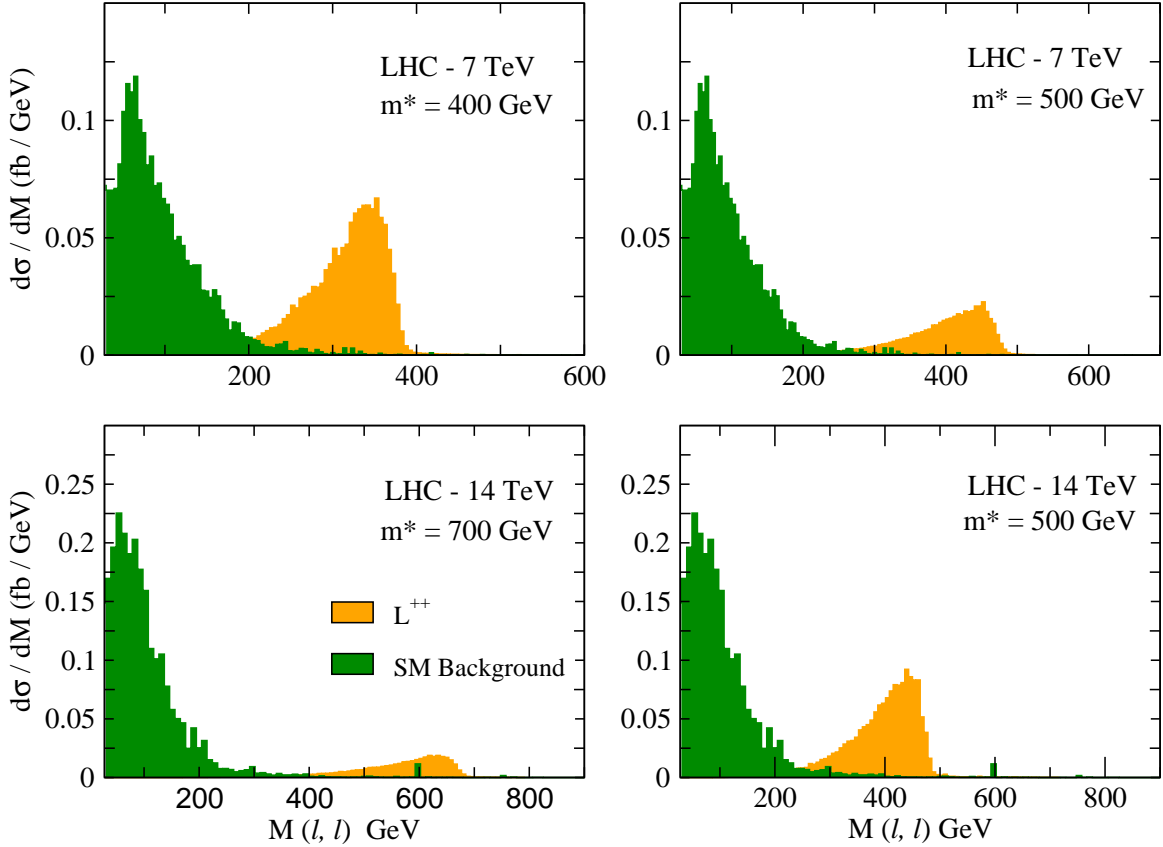}
\caption{\label{invariant_mass} (Color online) The invariant mass distribution for the same-sign-dilepton system  at $\sqrt{s}=7 $ TeV (top panels) and at $\sqrt{s}=14 $ Tev (bottom panels). The Standard Model background is with a dark filling (green online) and the excited lepton signal is with a light filling (orange online). The distributions in each plot are well separated, and the overlap between signal and background decreases with increasing values of $m^{*}$.}
\end{figure}

We have employed the following kinematic acceptance on the transverse momentum, rapidity and particle separation for the same final state particles:
\begin{eqnarray}
p_T (\ell)&=&15 \hbox{\ GeV} \: , \: |\eta (\ell)|<2.5 \\
p_T (\nu)&=&25 \hbox{\ GeV} \: , \: \Delta R(\ell^+,\ell^+)>0.5 
\end{eqnarray}
We have considered all the possible background processes provided by Standard Model for particle in the final state, in order to obtain a mass distribution for SSDL system in both cases, for the signal and for background events. The processes provided by Standard Model for the background are the following:
\begin{eqnarray}
p \, p \, \rightarrow \, W^{+} \, Z^{0} \, \rightarrow \, \ell^{-} \, \ell^{+} \, \ell^{+} \, \nu_{\ell}\\
p \, p \, \rightarrow \, W^{+} \, \gamma \, \rightarrow \, \ell^{-} \, \ell^{+} \, \ell^{+} \, \nu_{\ell}\\
p \, p \, \rightarrow \, \ell^{+} \, \nu_{\ell} \, \rightarrow \, \ell^{+} \, \gamma^{*} \, \nu_{\ell} \, \rightarrow \, \ell^{+} \, \ell^{+} \, \ell^{-} \, \nu_{\ell}
\end{eqnarray}
The results are showed in Fig.~\ref{invariant_mass}. The plots show the very small overlap between signal and background distributions. Thus we should be able to distinguish clearly the presence of doubly charged leptons despite the rather small signal cross sections.     


For a realistic study of the LHC phenomenology it is important to know at which integrated luminosity we could see some signal for excited doubly charged leptons, referring to the statistical significance. According to the previously shown mass distributions, we can choose a mass window below $m^{*}$ in order to calculate a cross section for the signal ($\sigma_{s}$) and a cross section for background ($\sigma_{b}$). In addition to this, we can set a relation between the statistical significance $s$, and integrated luminosity. Indeed from
\begin{equation}
\label{SS}
s=\frac{N_{s}}{\sqrt{N_{s}+N_{b}}}                       
\end{equation}
where $N_{s}$ is the number of signal events in the chosen invariant mass window:
\begin{equation}
\label{Nsig}
N_{s} = L \, \int^{m^{*}}_{m^{*}-\Delta m^{*}}\, dm_{(\ell^{+}, \ell^{+})} \frac{d\sigma_{s}}{dm_{(\ell^{+}, \ell^{+})}} = L \, \sigma_{s}                    
\end{equation}
and $N_{b}$ is the number of background events in the same invariant mass window:
\begin{equation}
\label{Nbg}
N_{b} = L \, \int^{m^{*}}_{m^{*}-\Delta m^{*}}\, dm_{(\ell^{+}, \ell^{+})} \frac{d\sigma_{b}}{dm_{(\ell^{+}, \ell^{+})}} = L \, \sigma_{b}                    
\end{equation}
Using the relations of Eqs.~\eqref{Nsig},\eqref{Nbg} into Eq.~\eqref{SS} one can easily solve for $L$:
\begin{equation}
L=s^{2}\left( \frac{\sigma_{s}+\sigma_{b}}{\sigma^{2}_{s}}  \right)                       
\end{equation}
In this way we can calculate the requested luminosity at LHC to see an excited doubly charged lepton, with a given mass, within a statistical significance $s=3$ (3 sigma effect) or $s=5$ (5 sigma effect). For a given mass window in going  from $\sqrt{s}=7~\text{TeV}$ to $\sqrt{s}=14~ \text{TeV}$ on can approximatively gain as much as an order of magnitude in the Luminosity $L$. According to the different features of the invariant mass distribution of the background and the signal we can distinguish the signal shape notwithstanding the small cross cross sections. 

\begin{figure}[t!]
\includegraphics[scale=1]{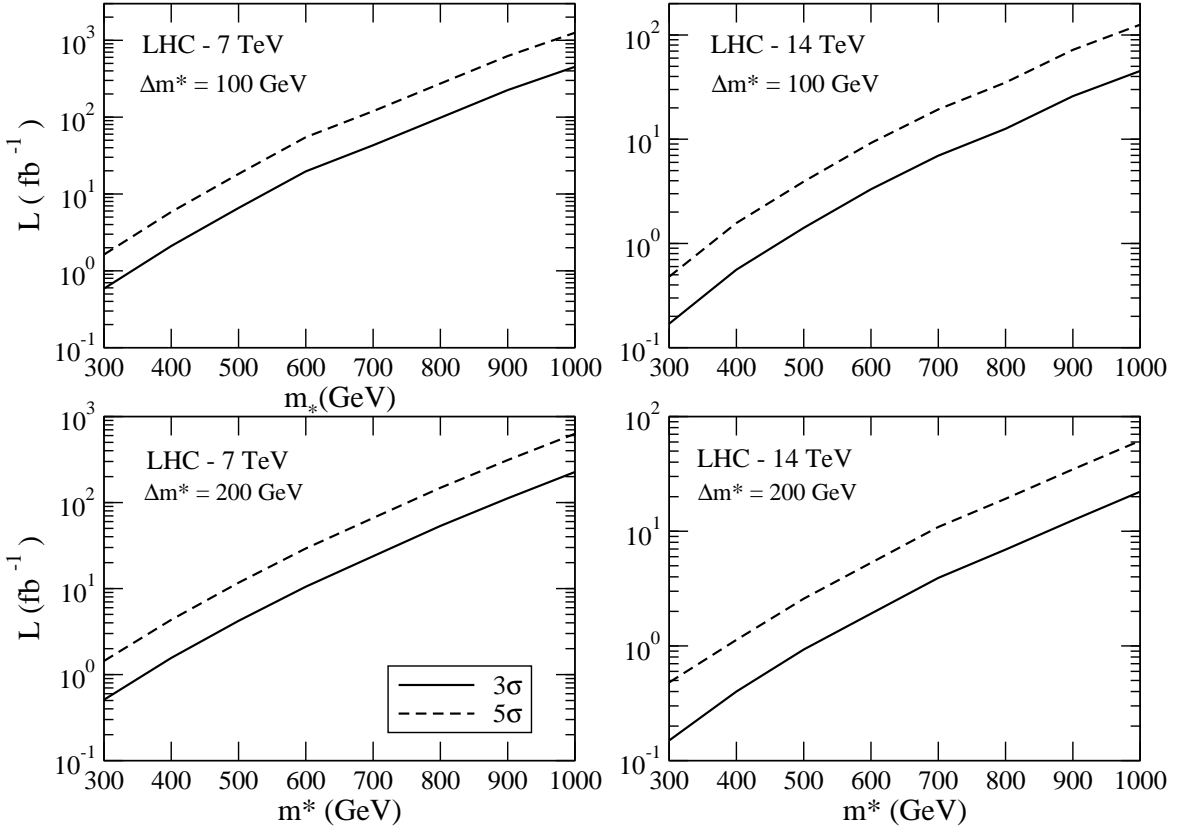}
\caption{\label{lumi_curves} The luminosity requested for observing an excited lepton with mass $m^{*}$ from 300 GeV up to 1 TeV. The two curves refer to a statistical significance equal to $3\sigma$ (solid line) and $5\sigma$ (dashed line), and for $\sqrt{s}=7$ TeV (left panels) and $\sqrt{s}=14$ TeV (right panels). In the top panels the signal and background invariant mass distributions have been integrated  in  an invariant mass window of $\Delta m^* = 100$ GeV, while for the bottom panels $\Delta m^* = 200$ GeV. }
\end{figure}

Fig.~\ref{lumi_curves} shows the curves of the  luminosity which is required to observe  at $\sqrt{s}=7~\text{TeV}$ (top left and bottom left panels) an excited lepton to a 3-sigma (5-sigma) level as function of $m^*$, the excited lepton mass. The signal and background have been integrated over two different mass windows: $\Delta m^* = 100$ GeV (top) and $\Delta m^* = 200$ GeV (bottom).  We see that the run at $\sqrt{s}=7$ TeV is sensitive at the $3$-sigma (5-sigma) level  up to a mass of order 600 GeV if $L=10^{-1} \, \text{fb}$ ($L=20^{-1} \, \text{fb}$). We also see (top right and bottom right panels) that the run at $\sqrt{s}=14~ \text{TeV}$  can reach a sensitivity at a 3-sigma (5-sigma) level up to $m^{*}=1000 \, \text{GeV}$ for $L=20^{-1} \, \text{fb}$ ($L=60^{-1} \, \text{fb}$).  

These conclusions are quite encouraging and prompted us to perform a preliminary study of the detector effects on our signature.

\section{Fast simulation and reconstructed objects}
\label{sec_fast+simulation}
The final step is to provide a more realistic description of our signature at the LHC. The distributions of the main kinematic variables given in the previous section are related to numerical CalcHEP outputs and they do not refer to some reconstructed objects. They are ideally detected with an efficiency of 100$\%$. The main difference is due to the effects of detector, which is characterized by an efficiency and a resolution in reconstructing kinematic variables $(E,P_{T})$. The consequence is the spreading of related distributions. 

Moreover in the previous sections we studied the signal and the background events as if those final state particles were the only ones to be produced in a proton-proton collision. For a fully realistic result, we must consider to add the hadronic activity which accompanies the production of the given final state of stable and color singlets particles. In order to achieve such a goal we interface  the CalcHEP output, given in a file following the Les Houches Accord Event (LHE)~\cite{lhe}  format, with the Pretty Good Simulator (PGS)~\cite{PGS4}. CalcHEP provides LHE output files, by its own event generator. This file contains the particles in the final state with their four-momentum before hadronization. This format is readable from Phytia \cite{PHYTHIA}, the MC generator used in this case just to provide a realistic description of the proton-proton initial interaction, a correct treatment of the beam remnants and evolve the final state particles in physics observables adding the showering and hadronization. PGS simulates the effect of a realistic detector reconstruction using a parametrization of the (resolution and efficiency) response. We use a parameterization compatible with one of the general purpose detectors actually taking data at the LHC (CMS or ATLAS)~\cite{MCsim}. We started with a sample of 1000 generated events for both signal and background. We consider the signal for an excited  lepton of mass $m^{*}=500$ GeV and the SM di-boson background, $WZ$ production, which is the main contribution expected~\cite{CMSNeu}. The strategy is the following:
\begin{enumerate}
\item study the kinematics of reconstructed leptons $\ell^{+} \, \ell^{+} \, \ell^{-} $, neglecting  the Missing Energy (MET) from the undetectable neutrino $\nu_{\ell}$;
\item request the presence of $\ell^{+} \, \ell^{+} \, \ell^{-} $ uniquely reconstructed by the detector (this is a PGS feature that increases the purity of the selection);
\item request the presence of at least one lepton with $P_{T}>50$ GeV, as this request ensures that the QCD background is largely suppressed;
\item get $m_{\left( \ell^{+} \, \ell^{+}\right)}$  distribution for signal and background on reconstructed objects with a certain hypothesis on the integrated luminosity. 
\end{enumerate}       

The general strategy outlined above applies equally well to all three families, but in PGS (as well as in a real detector) we will have different results, in terms of efficiency and resolution, requiring the reconstruction of electrons as opposed to muons or taus simply because different parts of the detector would be involved in  each case. \emph {In the following we  concentrate our analysis  to the electron case} ($\ell=e$).   

\begin{figure}[t!]
\begin{center}
\includegraphics[scale=0.425]{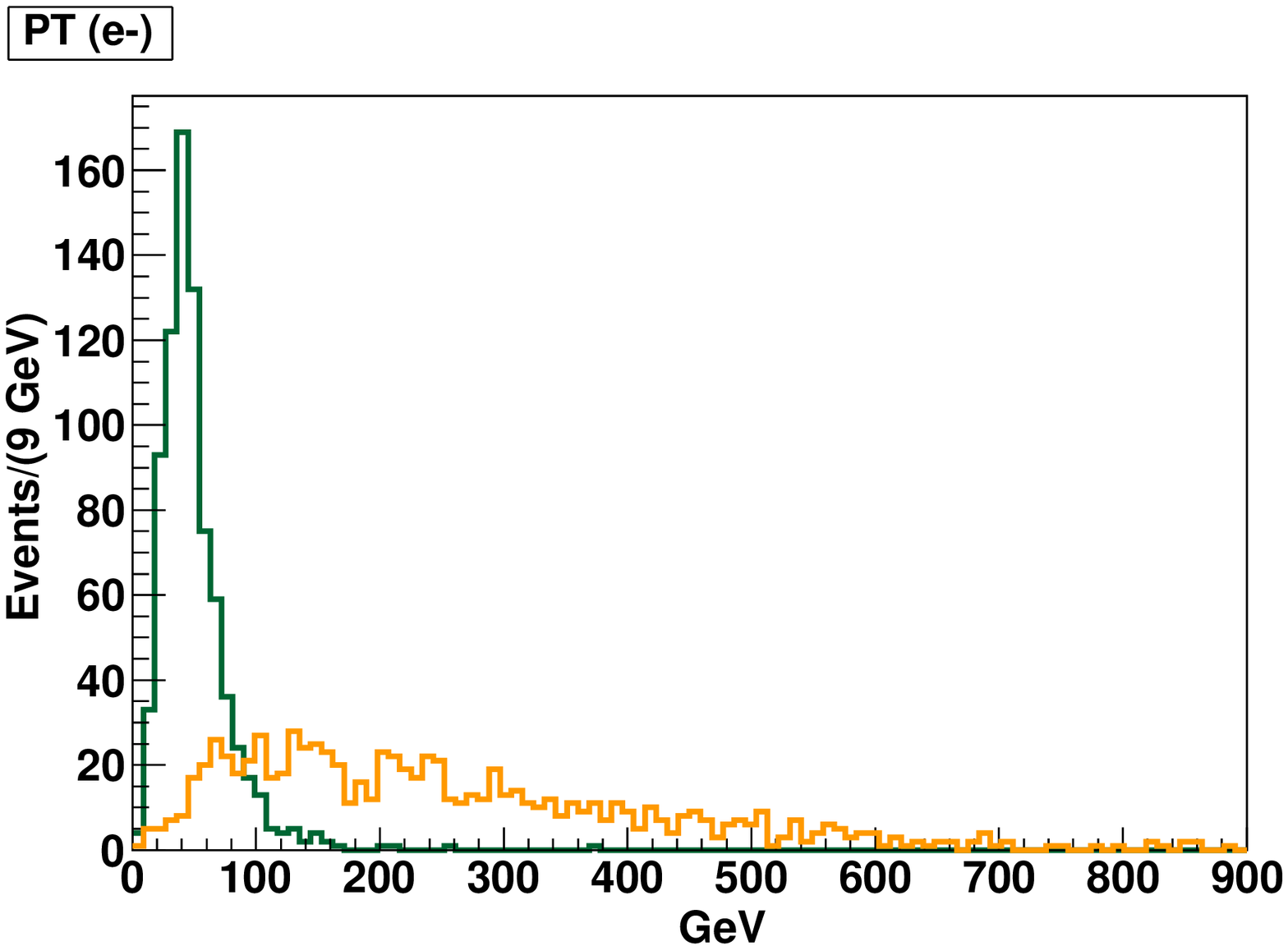}
\includegraphics[scale=0.425]{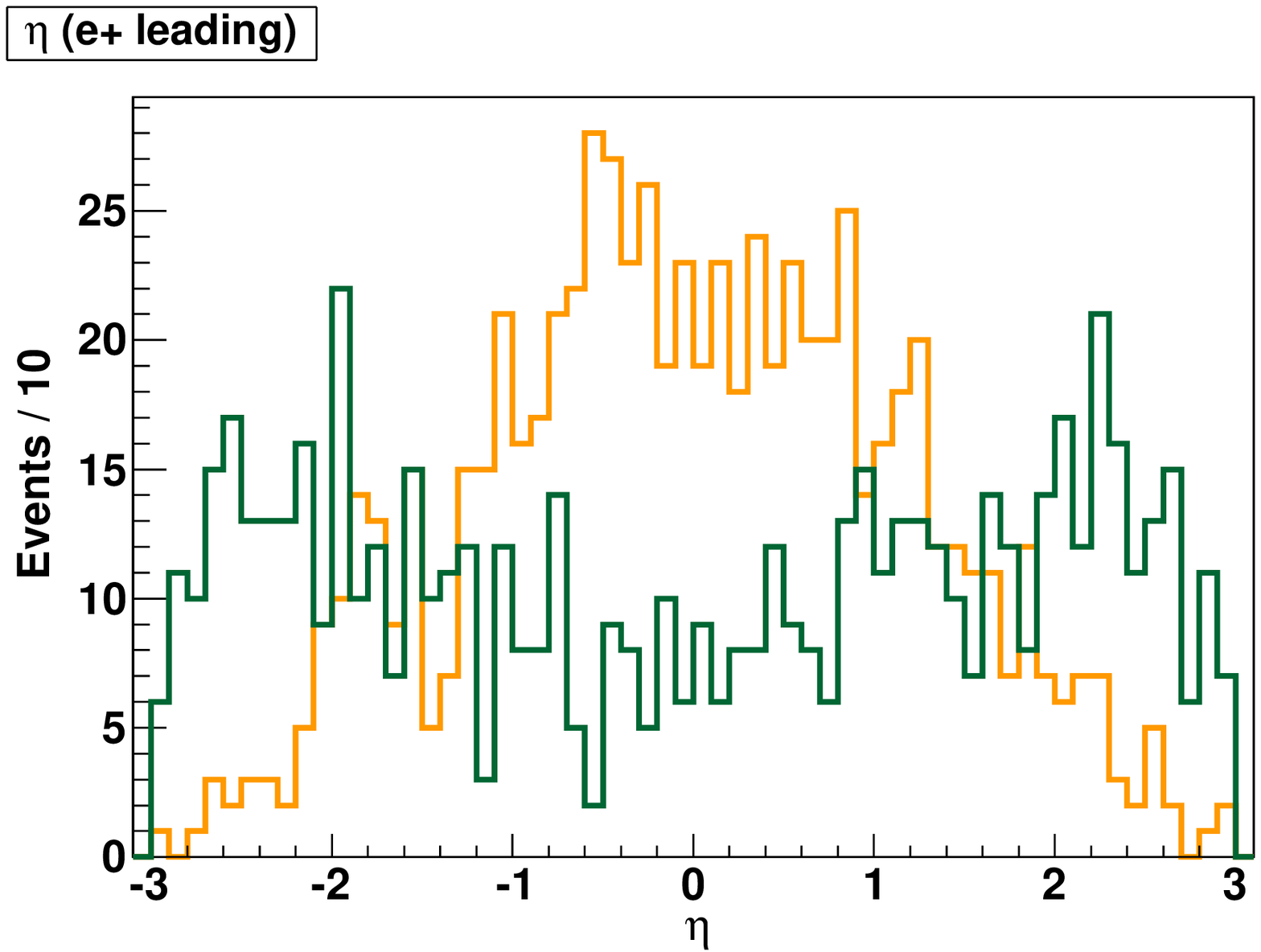}
\end{center}
\caption{\label{PTETA}(Color online) The main discriminating variables among background (dark line) and signal (light line) with $m_*=500$ GeV and $\sqrt{s}=7$ TeV. Left plot: the transverse momentum spectrum of the associated electron; Right plot: the pseudo-rapidity distribution of the $\ell^+$ with leading $p_T$. }
\end{figure}

Firstly we give a  kinematic characterization of the signal and background in terms of those variables that have the potential of discriminating better among the two. Fig.~\ref{PTETA} shows the transverse momentum spectrum (left) and the pseudorapidity distribution (right). Considering the kinematics, the experimental trigger has been simulated requiring to have a single  electron with $P_{T}<50 ~\text{GeV}$. The event selection cuts are done on the   $P_{T}$ of the negatively charged electron  $e^-$ and on the $\eta$ of the leading (most energetic) positively charged electron $e^+$.

In addition we impose the following main cuts: 
\begin{equation*}P_{T}\left( e^{-}\right) > 80~ \text{GeV} \quad , \quad |\eta \left( e_{1}^{+}\right) | <2 \end{equation*}
where $e_1^+$ is the  positron with the highest (leading) $P_T$.

Table~\ref{tabella} summarizes, for each request, the efficiency, for both signal and background, to get the expected number of  reconstructed final events.

\begin{table}[t!]
\caption{\label{tabella}The number of events for both signal and background after the applied cuts and selection criteria at $\sqrt{s}=7$ TeV.  The reconstruction efficiency and the event selection efficiency have been optimized with respect to the signal, \emph{assuming the electron channel}. The positron with the highest $P_T$ is denoted $e_1^+$.}
\begin{ruledtabular}
\begin{tabular}{c|c|c|c|c}
Events & SIG ($m*=500$ GeV) & BKG ($WZ$) & eff. SIG & eff. BGK\\
\hline
GENERATED EVENTS & 1000  & 1000 & 1 & 1 \\ 
\hline
$P_{T}\left( e\right) > 50$ GeV & 1000 & 760 & 1 & 0.76 \\
\hline
Reco $e^{+} \, e^{+} \, e^{-}$ & 650 & 403 & 0.65 & 0.53\\
\hline
$|\eta\left( e_{1}^{+}\right) |<2 \quad P_{T} \left( e^{-}\right)>80 $ GeV & 532 & 44 & 0.82 & 0.11 \\
\end{tabular}
\end{ruledtabular}
\end{table}

\begin{figure}[t!]
\centering
\includegraphics[scale=0.435]{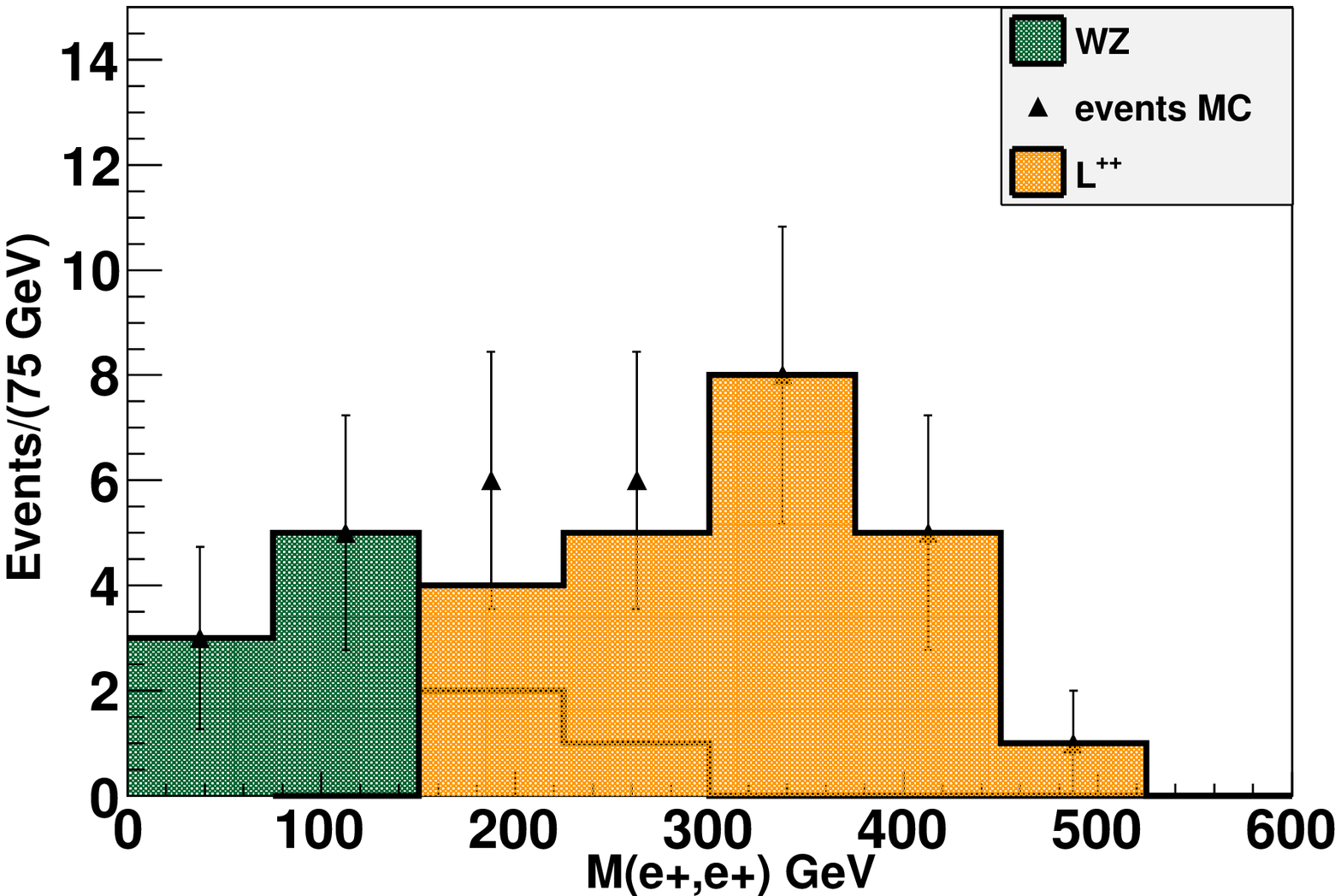}\!\!\!
\includegraphics[scale=0.435]{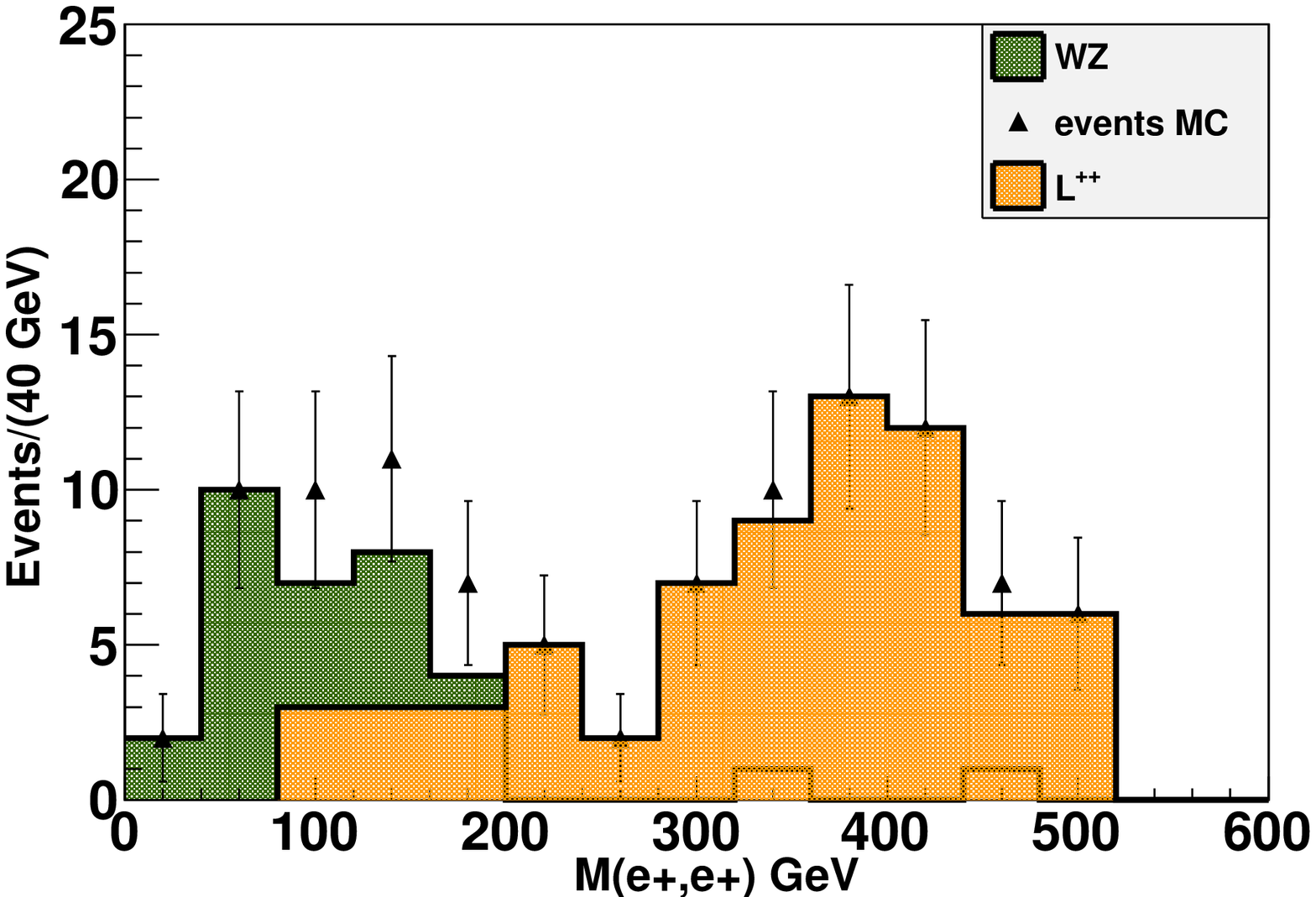}
\caption{\label{INV} (Color online) The invariant mass distribution of the SSDL after the fast detector simulation for the  run at $\sqrt{s}= 7$ TeV. A value of  $m_*= 500$ GeV  is assumed for the mass of the doubly charged lepton. The signal is the light filled line (orange online) while  the background is the dark filled line (green online).  Left panel is for an integrated luminosity $L=10$ fb$^{-1}$; Right panel is for $L=30$ fb$^{-1}$.}
\end{figure}

At $\sqrt{s}=7$ TeV, according to the previous efficiencies, one obtains $N_{S}=23$ and $N_{B}=11$ for an integrated luminosity of 10 $fb^{-1}$. If a higher luminosity of 30 fb$^{-1}$ could be obtained  then the expected number of events are  $N_{S}=70$ and $N_{B}=34$. The two plots in Fig.\ref{INV} show the \emph{reconstructed} invariant mass for both signal and background. We see clearly that  the separation of the distributions is largely preserved.

\section{Discussion and conclusions}
\label{sec_conclusions}

In this work we have discussed the phenomenology of exotic composite leptons of charge $Q=+2e$ at the LHC. Such states are allowed in composite models extended to include higher isospin multiplets, $I_{W}=1$ and $I_{W}=3/2$. Doubly charged leptons $L^{++}/L^{--}$ in these models couple with the Standard Model fermions electro-weakly only through the $W$ gauge boson. Thus  their decay channels are precisely identified.  These new resonant states are expected on general grounds if a further level of substructure exists.  

The model, originally discussed in ref.~\cite{YN} in its essential features, has been implemented in the CalcHEP software in order to be able to study the production cross section and kinematic distributions of the final state particles. The magnetic moment type interaction ($\sigma_{\mu\nu}$ coupling) is not present in CalcHEP and we implemented it making use of the mathematica package FeynRules.  Thus we were able to compare our analytical results with those of CalcHEP numerical sessions, such as parton cross sections and decay widths, in order to cross-check and validate the newly defined  CalcHEP model.

As regards the phenomenology of the doubly charged leptons we concentrated on the leptonic signature deriving from the cascade decays $ L^{++} \rightarrow W^{+} \ell^{+} \rightarrow \ell^{+} \, \ell^{+} \, \nu_{\ell}$ i.e. $p\,p \rightarrow \ell^{-} \left( \ell^{+} \, \ell^{+} \right) \, \nu_{\ell}$. Thus we focussed on a signature characterized by a low Standard Model background and a particular topology, that of \emph{like sign dileptons} which allows for a clear separation of the signal and background overcoming the difficulties of the rather low cross sections. 

We studied the main kinematic distributions for both signal and background, finding out clear differences among them. We showed that the invariant mass distribution of the same-sign dilepton system has a sharp end point corresponding to the excited lepton mass $m^{*}$. The same feature is absent in the invariant mass distribution of the SM background. Thus the invariant mass $m_{\left(\ell^{+}, \ell^{+} \right) }$ is the most discriminating variable among signal and background as shown in Fig.~\ref{invariant_mass}. We find that the $\sqrt{s}=7$ TeV run is sensitive at a 3-sigma (5-sigma) level to a mass of the order of 600 GeV if $L=10$ $fb^{-1}$ ($L=20$ $fb^{-1}$). The $\sqrt{s}=14$ TeV run can reach a sensitivity at a 3-sigma (5-sigma) level up to $m^{*}=1000$ GeV for $L=20$ $fb^{-1}$ ($L=60$ $fb^{-1}$). 

The parton level CalcHEP distributions are referred to ideal physical objects, as an ideal detector would reveal them, i.e. detection efficiency and misidentification of particles are not considered at all. In order to provide a more realistic description of our processes (signal and background) we developed an interface among the CalcHEP LHE output and general purpose detector simulator PGS~\cite{PGS4}. Thus the simulation of reconstruction is provided identifying some selection cuts that reject background events  as much as possible (low efficiency) and save signal events  as much as possible (high efficiency). The efficiency of the detector is considered: tracker resolution, calorimeter resolution and geometrical acceptance. The invariant mass distribution $m_{\left( {\ell^{+},\ell^{+}} \right) }$ is presented after the loss of events due to the detector efficiency and the smearing effect is included, as shown in Fig.~\ref{INV}.

We also discussed, see Section~\ref{mechanisms},  some other mechanisms of production of the exotic doubly charged leptons. In particular pair production, via Drell-Yan gluon-gluon fusion and contact interactions.  We find that even with the current strong lower bound on the contact interaction scale ($\Lambda_C>10$ TeV)  pair production via contact interactions is competitive at $\sqrt{s}=7$ TeV and even dominant at  $\sqrt{s}=14$ TeV, though leading to different signatures.  Single production of the exotic doubly charged leptons via contact interactions while certainly interesting and well motivated would require flavor conserving but non diagonal interactions whose analysis is demanded to a future work. In addition a comparative analyis of the contact interaction mechanism with the magnetic type gauge interactions would require to implement the contact interactions in CalcHEP. 

The final conclusion of this study is that the hypothetical doubly charged leptons peculiar of extended weak isospin composite models could be searched for at the LHC, and full fledged analysis by the experimental LHC collaborations could be started in the data samples already collected. 

\begin{acknowledgments}
This work is an outcome of the master thesis of S.~B. presented at the University of Perugia in December 2011.
The authors are indebted to M. Narain for having first suggested to study the phenomenology of the exotic states in the weak-isospin extended composite models.

The authors acknowledge constant interest, encouragement and support by the local Perugia CMS group.   
\end{acknowledgments}

\end{document}